\documentclass[universe,conferencereport,accept,moreauthors,pdftex,10pt,a4paper]{mdpi}
\pdfoutput=1

\usepackage{bm,booktabs,graphicx,microtype,multirow,setspace,soul}

\firstpage{1}
\articlenumber{14}
\doinum{10.3390/universe4010014}
\pubvolume{4}
\pubyear{2018}
\copyrightyear{2018}
\history{Received: 30 November 2017; Accepted: 16 January 2018; Published: 23 January 2018}

\newcommand\ti[1]{{\mbox{\scriptsize #1}}} 
\newcommand{\beq}{\begin{equation}} 
\newcommand{\eeq}{\end{equation}} 
\newcommand{\beqa}{\begin{eqnarray}} 
\newcommand{\eeqa}{\end{eqnarray}} 

\Title{From Heavy-Ion Collisions to Compact Stars: Equation of State and Relevance of the System Size}
\Author{Sylvain~Mogliacci *\orcidA{}, Isobel~Kolb\'e and W.~A.~Horowitz \orcidC{}}

\AuthorNames{Sylvain~Mogliacci, Isobel~Kolb\'e and W.~A.~Horowitz}

\address[1]{Department of Physics, University of Cape Town, Rondebosch 7701, South Africa;%
\linebreak isobel.kolbe@gmail.com (I.K.); wa.horowitz@uct.ac.za (W.A.H.)}%
\corres{Correspondence: sylvain.mogliacci@uct.ac.za}

\abstract{
In this article, we start by presenting state-of-the-art methods allowing us to compute moments related to the globally conserved baryon number, by means of first principle resummed perturbative frameworks. We focus on such quantities for they convey important properties of the finite temperature and density equation of state, being particularly sensitive to changes in the degrees of freedom across the quark-hadron phase transition. We thus present various number susceptibilities along with the corresponding results as obtained by lattice quantum chromodynamics collaborations, and comment on their comparison. Next, omitting the importance of coupling corrections and considering a zero-density toy model for the sake of argument, we focus on corrections due to the small size of heavy-ion collision systems, by means of spatial compactifications. Briefly motivating the relevance of finite size effects in heavy-ion physics, in opposition to the compact star physics, \mbox{we present} a few preliminary thermodynamic results together with the speed of sound for certain finite size relativistic quantum systems at very high temperature.}


\begin{document}


\section{Introduction}\label{sec:Intro}
The weak coupling expansion of the Quantum Chromodynamics (QCD) grand potential density (which we further call ``free energy density'', since both are equivalent in the limit of zero density in infinite volume systems), is known to be a central quantity for the thermodynamics of a deconfined system such as those created in Heavy-Ion Collisions (HIC). If computed naively, there are however certain details that need to be improved, and we are going to discuss them in the following.

First of all, such a weak coupling expansion appears not to converge at phenomenologically moderate temperatures~\cite{Convergence&WeakCoupling_Blaizot}, relevant to the quark-hadron phase transition which takes place around the pseudo-critical temperature $T_\ti{c}=154\pm 9$ MeV for vanishing quark chemical potentials $\mu_f$~\cite{Peikert_Tc,Bazavov_Tc}. Lattice Monte Carlo simulations cannot be used when the chemical potentials, and more specifically the baryon chemical potential, are non-zero due to the so-called sign problem of QCD~\cite{deForcrand:2010ys,Gupta:2011ma}. Consequently, a lot of effort has been put in developing frameworks allowing access to reliable results, from first principles, by continuing the known perturbative outcome toward the non weakly coupled phase transition region, including at non-zero density~\cite{Blaizot:2001vr,Andersen:2012wr,Mogliacci:2013mca,Haque:2014rua}. In the light of the current~\cite{RHIC,LHC} and forthcoming~\cite{FAIR,NICA} experiments, the non perturbative lattice simulations together with these resummation frameworks allow for further insights in the study of the QCD phase diagram~\cite{Mogliacci:2013mca,Mogliacci:2013iwa}.

Then, a different issue which unfortunately lacks more thorough investigations, is related to the fact that the deconfined systems which are briefly created in collider experiments have finite, and in fact comparatively rather small volumes. For instance, their characteristic sizes are at best of the order of $R\sim$ 4--8~Fermis for lead--lead collisions at $\sqrt{s}=2.76$ TeV~\cite{Aamodt:2011mr}, and $R\sim$ 1--2~Fermis for proton--lead collisions at $\sqrt{s}=5.02$ TeV~\cite{Adam:2015pya}. It is then trivial to compare these lengths to a temperature $T$ which is typical, say $T\sim 2 T_\ti{c}$ to be conservative. Moreover, by doing so, one can see that the relevant dimensionless parameter, namely $\Delta = T\!\times\!R$, falls into neither of the extreme regimes: It indeed ranges in between a few units to roughly ten (it becomes, however, nearly an order of magnitude smaller if considering an hydrodynamic cell in local equilibrium). For comparison, systems relevant to the description of a compact star are in a complete different regime, with $\Delta=\infty$ being an excellent approximation (accounting for a small, yet non-zero, temperature~\cite{Vuorinen:2016pwk}, or evaluating the dimensionless parameter $\mu\!\times\!R$ instead). Thus, it appears that in the context of a quark--gluon plasma relevant to the heavy-ion physics, an important question is: Which physical quantities happen to be sensitive to the finite size of the system, and which are not? In this article, we will first briefly recall certain aspects of the QCD thermodynamics for infinite size systems, together with the corresponding fluctuations of globally conserved charges. Next, still relevant to infinite size systems, we will present two state-of-the-art frameworks for resumming the weak coupling expansion of QCD at finite temperature and density. Then, we shall move on to briefly introduce the proposed framework for finite size corrections. After which we will start reviewing results corresponding to the fluctuations and correlations of conserved charges, comparing the resummed perturbative framework results to those of lattice QCD collaborations. Finally, we will present a few preliminary results concerning finite size corrections, using a single non interacting massless scalar field at zero chemical potential as a toy model. A certain number of points will not be emphasized, and we refer the readers to~\cite{Andersen:2012wr,Mogliacci:2013mca,Mogliacci:2014pxa} and references therein for further details on the frameworks as well as all the results for finite density investigations, and to~\cite{To_appear} for more details on the finite size preliminary results.


\section{Charge Fluctuations in Infinite Size Systems}\label{sec:Thermo&Fluctu_QCD}

We start by briefly recalling the link between the Hamiltonian of Quantum Chromodynamics $H_\ti{QCD}$ and its partition function $Z_\ti{QCD}$, which in infinite size systems reads:
\begin{equation}
Z_\ti{QCD}\left(T,\mu_f;V\right) \equiv \mbox{Tr\ }\exp\left[-\frac{1}{T}\Bigg(H_\ti{QCD}-%
\sum_f \mu_f\ Q_f\Bigg)\right] = \mbox{Tr\ }\Big(\rho_\ti{QCD}\Big) \label{PartitionFunction} ,
\end{equation}
where $Q_f$ and $\mu_f$ respectively denote the conserved charges and the corresponding chemical potentials. Furthermore, in the following, $\left\langle\vartheta\right\rangle\equiv\mbox{Tr\ }\left(\vartheta\cdot\rho_\ti{QCD}\right)/Z_\ti{QCD}$ will denote thermal averages.
While we mainly consider the up, down, and strange quarks with respective chemical potentials $\mu_\ti{u}$, $\mu_\ti{d}$, and $\mu_\ti{s}$, one can also express the partition function in terms of the baryon charge, the electric charge, and the strangeness conserved number, with ($\mu_\ti{B},\ \mu_\ti{Q},\ \mu_\ti{S}$) instead. From Equation~(\ref{PartitionFunction}), one can see that the mean and (co)variance of two conserved charges can be expressed in terms of derivatives with respect to the chemical potentials, following:
\beqa
\left\langle Q_f \right\rangle&=&T\,\frac{\partial}{\partial \mu_f} \log Z_\ti{QCD} , \label{Mean} \\
\left\langle \left(Q_f-\left\langle Q_f \right\rangle\right) \cdot \left(Q_g-\left\langle Q_g \right\rangle\right) %
\right\rangle&=&T^2\,\frac{\partial^2}{\partial \mu_f \partial \mu_g} \log Z_\ti{QCD} , \label{Covariance}
\eeqa
which is straightforwardly related to the first and second order cumulants, respectively. These above quantities, referred to as susceptibilities, are defined for the quark numbers by:
\beq
\chi_\ti{u$_i$\,d$_j$\,s$_k$\, ...}\left(T,\left\{\mu_f\right\}\right) \equiv %
\frac{\partial^n p_\ti{QCD}\left(T,\left\{\mu_f\right\}\right)}%
{\partial\mu_\ti{u}^i\, \partial\mu_\ti{d}^j \, \partial\mu_\ti{s}^k\, ...} ,
\eeq
with $n=i+j+k+...$, and where we recall that equilibrium thermodynamic quantities such as the pressure follow simple relations in infinite size systems like:
\beq
p_\ti{QCD} = \frac{\partial \left(T \log Z_\ti{QCD}\right)}{\partial V}\ %
\xrightarrow[V\rightarrow\infty]{ }\ \frac{T}{V}\,\log Z_\ti{QCD} .
\eeq

It should be noted that one may also consider any other conserved charge, instead of the \mbox{quark numbers}. Furthermore, the number susceptibilities are, in general, important as they give information on the correlations and fluctuations of the globally conserved quantum numbers. Therefore, they turn out to be very practical probes for the changes of degrees of freedom across the transition region, specifically when the conserved charge is the baryon number. They are also directly related to the corresponding cumulants, and thus provide some crucial information about the probability distribution of the baryonic degree of freedom, together with insights on the existence and location of a possible critical point on the QCD phase diagram. For more detail on the use of conserved charge cumulants in HIC, we refer the readers to~\cite{Satz_HIC,Koch_Cumulants}.

\section{Resummed Perturbative Quantum Chromodynamics in Infinite Size Systems}\label{sec:Framework_InfiniteSize}
\vspace{-6pt}
\subsection{Resummation Inspired from Dimensional Reduction}

The Dimensional Reduction (DR) phenomenon at asymptotically high temperature, which can generally be understood as the appearance of certain effective degrees of freedom in a lower dimension, is well known to account properly for the dynamics of energy scales up to the order $gT$ in QCD~\cite{DimRedPheno1,DimRedPheno2}. Such a procedure is carried out using an effective field theory which is called Electrostatic QCD (EQCD), by integrating out the hard degrees of freedom followed by a careful matching of the effective theory with the original one. EQCD is a three-dimensional SU($N_\ti{c}$) Yang-Mills theory coupled to an adjoint Higgs field~\cite{BraatenNieto_EQCD,Kajantie_EQCD}, which accounts for all the infrared divergences encountered in the weak coupling expansions~\cite{Linde_IR_Pb}. And as such, EQCD provides a rigorous framework for carrying out higher order loop computations in high temperature perturbative QCD.

Using this knowledge, one can rewrite the QCD pressure as:
\beq
p_\ti{QCD} = p_\ti{hard}(\mbox{\small$g$}) + T\, p_\ti{EQCD}%
(\mbox{\small$m_\ti{E},g_\ti{E},\lambda_\ti{E},\zeta$}) ,
\eeq
where the parameters $m_\ti{E}(g),g_\ti{E}(g),\lambda_\ti{E}(g)$ and $\zeta(g)$ are also functions of the temperature and the quark chemical potentials, and admit expansions in powers of the four-dimensional gauge \mbox{coupling $g$}. \mbox{The contribution} $p_\ti{hard}$ is relevant to the hard scale ($\sim$$T$), and can be computed through a direct loop expansion in QCD. The contribution $p_\ti{EQCD}$ is, on the other hand, relevant to the soft ($\sim$$gT$) and ultrasoft ($\sim$$g^2T$) scales. This contribution is accessible from the partition function of EQCD, and through a partial four-loop order even accessible by means of loop expansion only.

In principle, when computing the EQCD pressure in order to be able to access the full QCD pressure, the entire result (with the EQCD parameters) must be Taylor expanded in powers of $g$ around small values. However, it was suggested in~\cite{Convergence&WeakCoupling_Blaizot} and then first applied at zero chemical potential in~\cite{Mikko&York_Quark_Mass}, that one can simply consider both $p_\ti{hard}$ and $p_\ti{EQCD}$ as functions of the EQCD parameters. Doing so, and not re-expanding them in powers of $g$ resums certain higher order contributions while keeping all correct contributions up to and including the order $g^6 \log(g)$~\cite{Keijo&Mikko&York,Aleksi_Pressure}. As a byproduct, the theoretical uncertainties through the renormalization scale dependence of the result is substantially reduced, \mbox{and the} convergence properties are thereby improved.

\subsection{Hard-Thermal-Loop Perturbation Theory}

The use of a variationally improved perturbation theory framework has been known for decades to allow important higher order resummations as well~\cite{Opt_Kneur,Opt_Peter_SPT}. The introduction of a certain relevant term to be added and subtracted from the Lagrangian density, allows for the treatment of the added/subtracted piece with the non-interacting/interacting part. By doing so, one actually interpolates between the original theory and a theory having appropriately dressed propagators and vertices, while recovering the original theory in the end by setting (see below) $\delta = 1$. For QCD, \mbox{the procedure} is of course more complicated given gauge invariance, and the relevant term is the non-local hard-thermal-loop effective action~\cite{Frenkel&Taylor_HTL,Braaten&Pisarski_HTL}. This procedure is called Hard-Thermal-Loop perturbation theory (HTLpt)~\cite{Jens&Mike&BraatenHTLpt,Jens&Mike&BraatenHTLptbis}, and the subsequent Lagrangian reorganization reads:
\beq
{\cal L_{\rm HTLpt}}=\Big[{\cal L}_{\rm QCD}+(1-\delta)\ {\cal L}_{\rm HTL}\Big]%
\Big|_{g\rightarrow\sqrt{\delta}g}+\Delta{\cal L}_{\rm HTL} ,
\eeq
where ${\cal L}_{\rm HTL}$ is the gauge invariant HTL improvement term, and $\delta$ a formal expansion parameter set to one after the expansion. We notice, in the above, that $\Delta{\cal L}_{\rm HTL}$ is a counter term necessary to cancel all the ultraviolet divergences introduced by the reorganization of the perturbative series, from the ground state of an ideal gas of massless particles to the ground state of an ideal gas of massive quasiparticles.


\section{On the Finite Size Corrections}\label{sec:Finite_Size_Corrections}

First of all, we would like to refer the readers to the forthcoming article~\cite{To_appear}, where all the details from the conceptual to computational aspects will be exposed to a greater extent.

We wish now to drastically simplify the approach to the quark--gluon plasma created in high energy collisions, in order to make a first step in accounting for its finite size. To this end, we will disregard the importance of accounting for the interaction, and consider a zero-density toy model for a start: We chose a single non-interacting massless scalar field at zero chemical potential. %
And indeed, \mbox{it is} worth noticing that such a massless scalar field is, in fact, quite insightful, for it contains information relevant to a gas of non-interacting gluons---albeit a group theory prefactor (to the free energy), which is not present in any of the displayed quantities here, since we conveniently show only the appropriate ratios. %
We will then solely focus on the corrections due to the small size of our system.
Furthermore, we are not discussing here a real finite volume system, leaving it for~\cite{To_appear} and referring to works such as~\cite{Karsch:2015zna} for other types of finite volume investigations (this time, relevant to \mbox{hadronic systems}).

Instead, we will simply discuss a quantum relativistic system whose dynamics are governed by the aforementioned field theory, coupled to a heat bath at temperature $T$ and geometrically confined in between two infinite parallel planes separated by a distance $L$. More precisely, motivated by physical arguments~\cite{To_appear}, we undertake a spatial compactification by assuming Dirichlet boundary conditions on both infinite planes, and finally arrive, explicitly, to the free-energy density of a neutral, massless, and non-interacting scalar field, which reads:
\beqa
f(T,L)&\equiv& F(T,L,A)/V\nonumber\\%
&=& -\frac{\pi^2T^4}{90}+\frac{\zeta(3) T^3}{4\pi L}-\frac{\zeta(3) T}{16\pi L^3}\nonumber\\%
&&-\frac{T^2}{8 L^2}\times\sum_{s=1}^{+\infty}\bigg[\frac{\text{csch}^{2}\left(2\pi TL\times s\right)}{s^{2}}\bigg]%
-\frac{T}{16\pi L^3}\times\sum_{s=1}^{+\infty}\bigg[\frac{\coth\left(2\pi TL\times s\right)-1}{s^{3}}\bigg] , \label{FreeEnergyDensity}
\eeqa
and where $F(T,L,A)$ is the free energy, $A\equiv V/L$ being the area of each of the infinite parallel planes. A few comments are now in order, concerning the Equation~(\ref{FreeEnergyDensity}). The above expression is an exact analytic representation of the free energy density of our system coupled to the heat bath and in between the parallel planes. Moreover, it is resummed to be exponentially fast in terms of convergence for practical numerical evaluations (when the sums are then truncated; see~\cite{To_appear} for more detail). Finally, let us notice that the first term in the first line of~(\ref{FreeEnergyDensity}) is the usual (fully) thermal non-compactified result. Next to it, the two simple terms are part of the thermal corrections to the Casimir (geometric) effect due to the presence of boundaries inside the heat bath. The last two terms, each containing an infinite summation, account for both the rest of the thermal corrections to the geometric effect and for the so-called zero-temperature Casimir result. Indeed, it can be checked that applying the (well defined) limit $T\rightarrow 0$ to the above will give us:
\beq
f(T=0,L)=-\frac{\pi^2}{1440~L^4} , \label{FreeEnergyDensityZeroTemp}
\eeq
which is responsible for the well-known zero-temperature Casimir pressure $p_{\rm Cas}\equiv -\pi^2/(480~L^4)$. Thus, our expression is not only a very compact one which converges exponentially fast when the sums are truncated for nearly any values of $T$ and $L$, but it rightfully reproduces the two appropriate limits, namely $L=\infty$ and $T=0$ respectively.

Let us now present all the results, some concerning the (QCD) infinite volume case at finite density and some being relevant to the (toy model) finite size case at zero density.


\section{Results and Discussion}\label{sec:Results&Discussion}

In the present section, we refer the readers to~\cite{Mogliacci:2013mca} for more detail on the setting of the parameters in the case of the finite density results, the bands corresponding to conservative variations of all the resummed perturbative parameters. We also refer to~\cite{To_appear} for a forthcoming thorough investigation concerning the finite size preliminary results which we are presenting. Concerning the finite density results, the blue band corresponds to the DR result while the red and orange bands are the exact one-loop and truncated three-loop HTLpt results. The dashed curves inside the bands correspond to the central values of the renormalization and QCD scales. As for the finite size preliminary results, \mbox{all quantities} which are presented here are relevant to a system in between two infinite parallel planes separated by a distance $L$, and in contact with a heat bath at temperature $T$.

\subsection{Quantum Chromodynamics Infinite Volume Case at Finite Density}\vspace{-6pt}
\subsubsection{Low Order Susceptibilities}

\begin{figure}[H]\centering\includegraphics[scale=0.325]{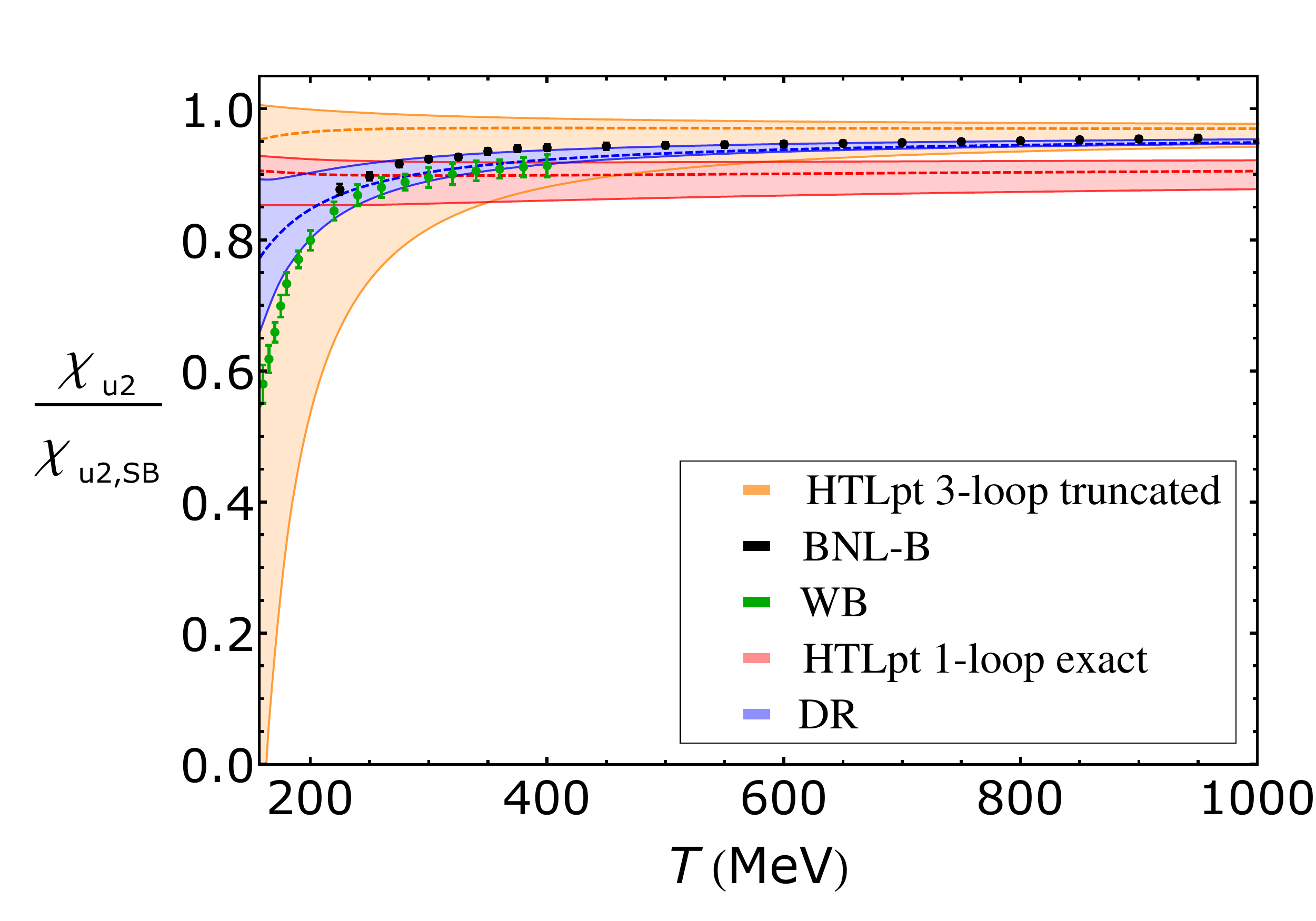}
\caption{Second-order diagonal quark number susceptibility, normalized to its non-interacting limit. The truncated three-loop HTLpt result is from~\cite{HTLpt_Finite_MU_3Loop1} and the lattice data from the BNL--Bielefeld~\cite{Bielefeld_Lattice_HighT} (BNL--B) as well as from the Wuppertal--Budapest~\cite{WB_Lattice1} (WB) collaborations.}\label{chiu2_Nf3}
\end{figure}

\begin{figure}[H]\centering\includegraphics[scale=0.30]{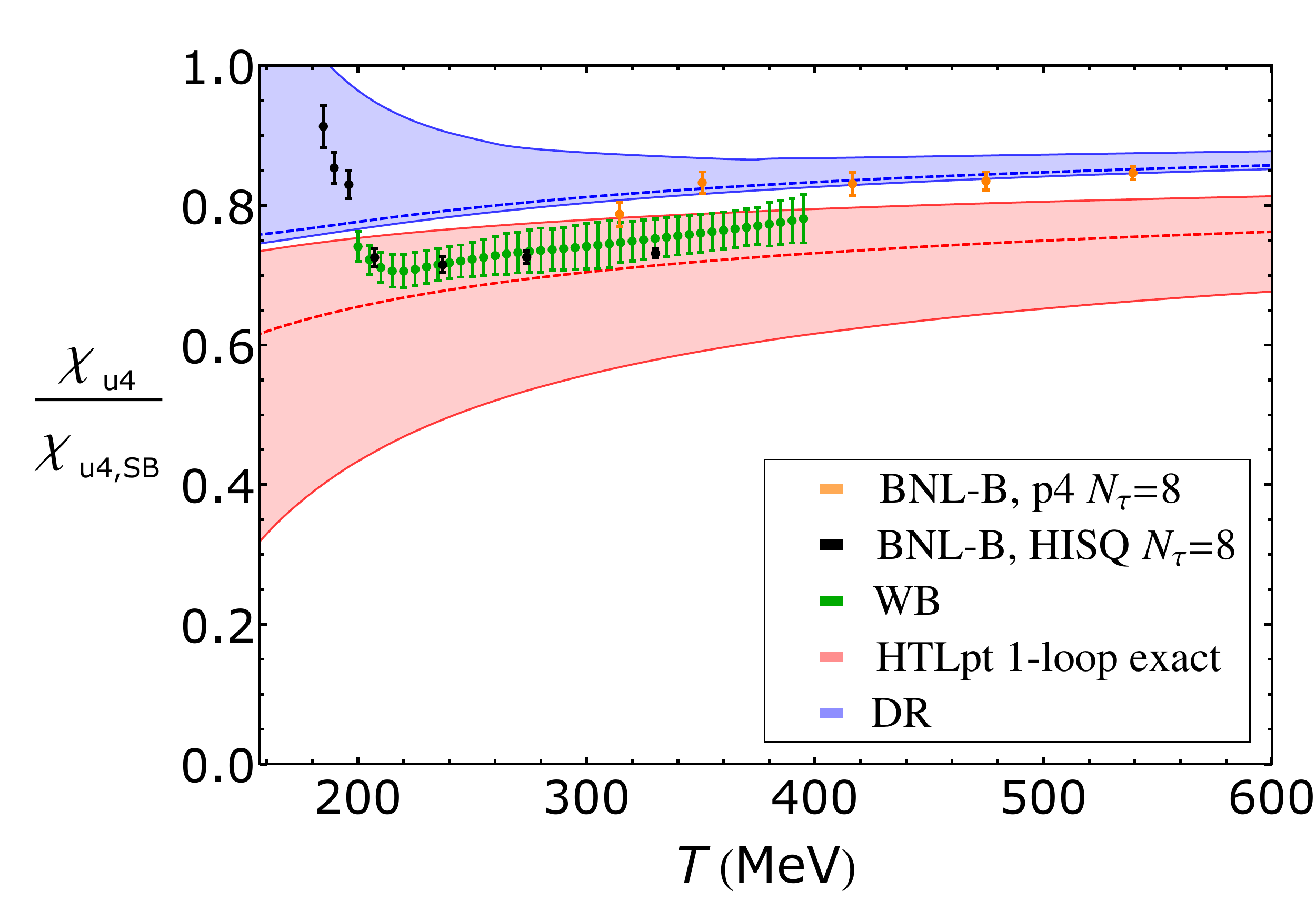}\!\!\!\!\includegraphics[scale=0.30]{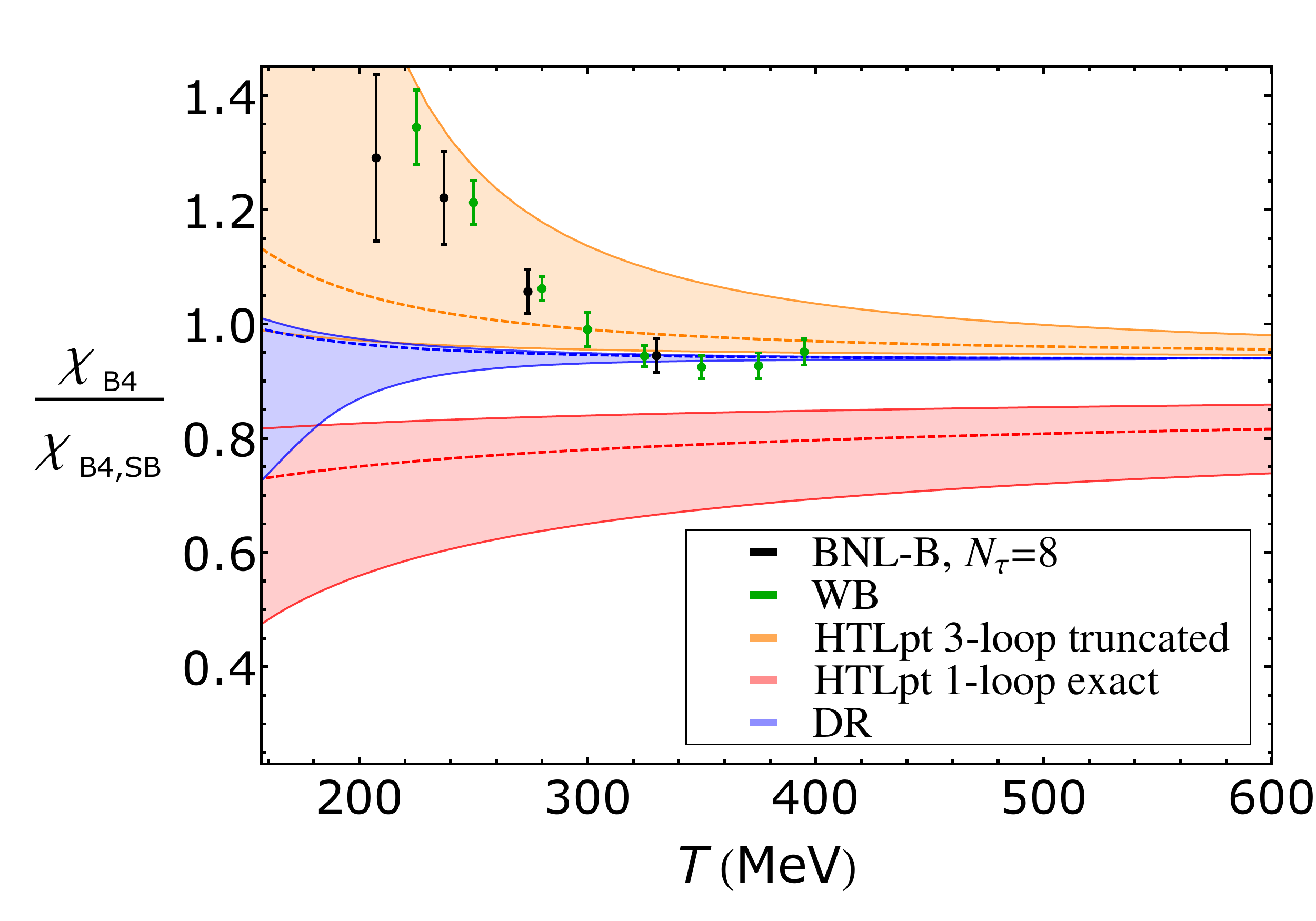}
\caption{Fourth-order diagonal quark (\textbf{left}) and baryon (\textbf{right}) number susceptibilities, normalized to the non-interacting limits. The truncated three-loop HTLpt result is from~\cite{HTLpt_Finite_MU_3Loop1} and the lattice data from the BNL--Bielefeld~\cite{Bielefeld_Lattice1,Bielefeld_Lattice2,Bazavov_Lattice} (BNL--B) as well as the Wuppertal-Budapest~\cite{Wup_Bud_Chi4&Ratio} (WB) collaborations.}\label{chiu4_Nf3&chiB4_Nf3}
\end{figure}

First, we display low order quark and baryon number susceptibilities in Figures~\ref{chiu2_Nf3} and~\ref{chiu4_Nf3&chiB4_Nf3} (left), and in Figure~\ref{chiu4_Nf3&chiB4_Nf3} (right), respectively. The second- and fourth-order diagonal number susceptibilities are normalized to their non-interacting limits.

\vspace{-12pt}

From the width of the bands, we clearly see that the DR scale dependence is extremely small for perturbatively relevant temperatures. Moreover, both DR and HTLpt are in accordance with each other, while agreeing quite well with the non perturbative lattice results down to $T\sim$ 200--400 MeV.

\subsubsection{Kurtoses}

Next, we are presenting the kurtosis, a certain ratio of the fourth and second order quark or baryon susceptibilities. It is a measure of how strongly peaked a quantity is, most often used to measure how a critical point is approached during a phase transition~\cite{Stephanov_Kurtosis}.

\begin{figure}[H]\centering\includegraphics[scale=0.30]{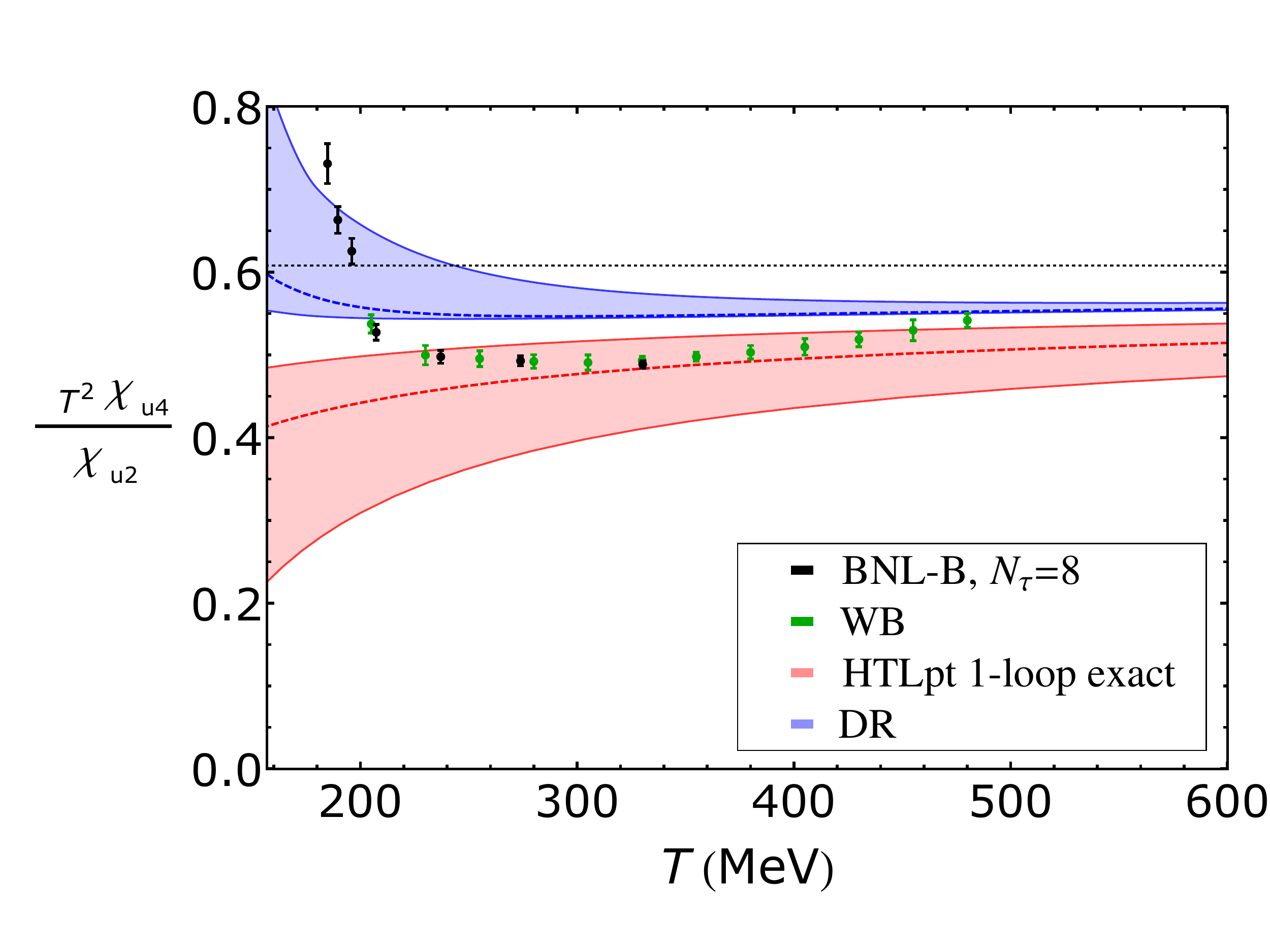}\!\!\!\!\includegraphics[scale=0.30]{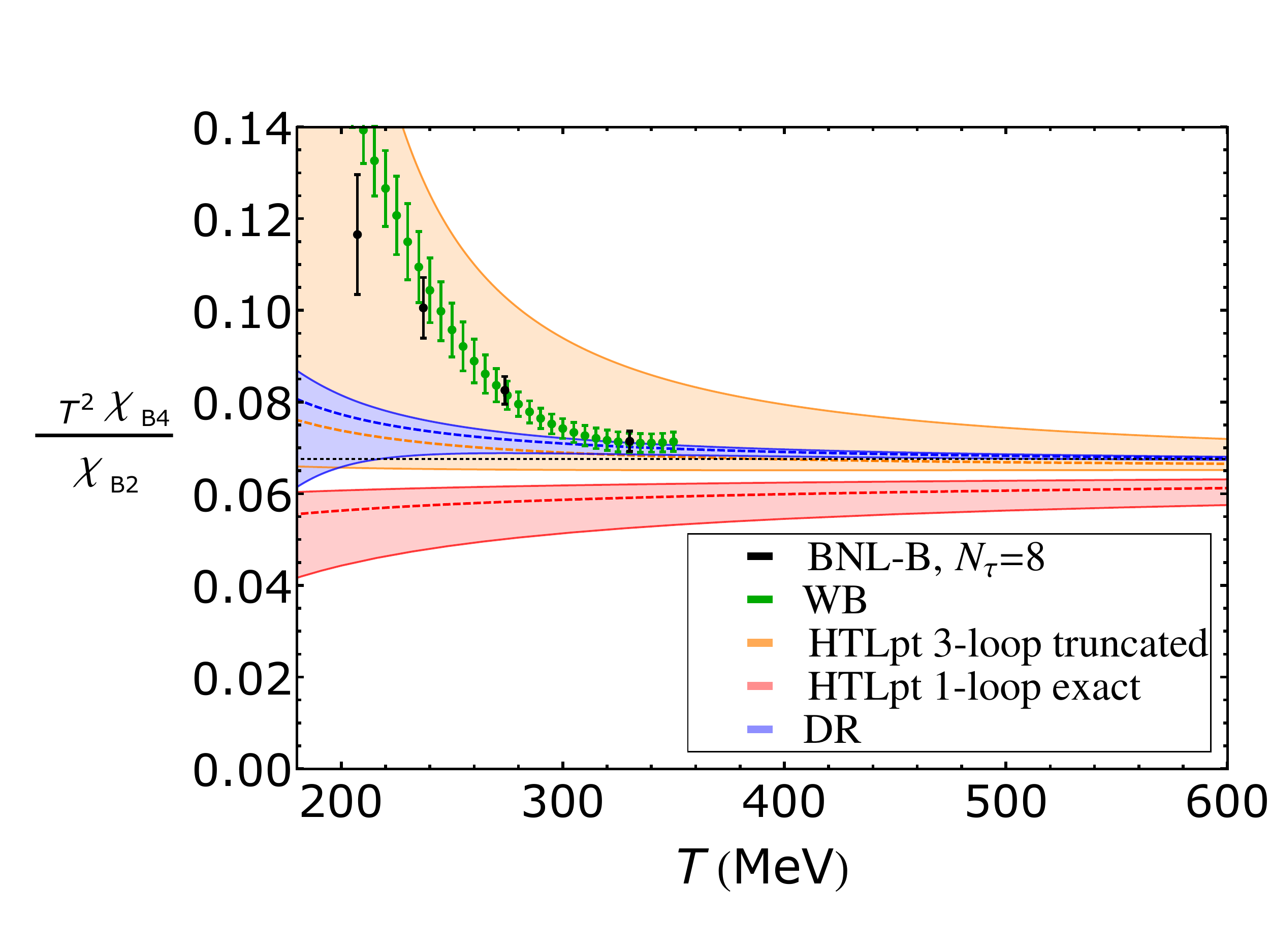}
\caption{Ratios of low order susceptibilities for the quark (\textbf{left}) and baryon (\textbf{right}) numbers. \mbox{The lattice} data is from~\cite{Bielefeld_Lattice1,Bielefeld_Lattice2} (BNL--B) and~\cite{Wup_Bud_Chi4&Ratio,Ratio_Wupp_Bud_Baryon} (WB). The three-loop HTLpt result is obtained from the corresponding cumulants of~\cite{HTLpt_Finite_MU_3Loop1}. The black dashed (straight) lines denote the Stefan Boltzmann limits.}\label{ratios}
\end{figure}

In Figure~\ref{ratios} (left), we plot the DR and HTLpt results together with lattice data which seem to agree with the one-loop HTLpt band at temperatures of $T \sim$ 300--400 MeV, however approaching the DR prediction at higher temperatures. The latter reproduces the overall trend of the lattice data better. On the right hand figure, both the three-loop HTLpt and DR predictions seem to agree with the lattice data at around $T \sim 350$ MeV, albeit the DR prediction is much more predictive. Both resummed perturbative results converge to the Stefan Boltzmann limit faster than for the result relevant to quark numbers. This tends to comfort the expectation that the medium should be less sensitive to the hadronic degrees of freedom in this range of temperatures.

\vspace{-12pt}

\subsection{Toy Model Finite Size Case at Zero Density}
\vspace{-6pt}
\subsubsection{Finite Size Corrections to the Thermodynamics}

We start by plotting the ratio of the free energy with its non-compactified limit (i.e., for which $L\rightarrow\infty$). Even though the intrinsic asymmetry of such a finite size system implies that the actual pressure along the planes may not be the same as the pressure across them, we recall~\cite{To_appear} that in the non-compactified limit, both pressures reduce to minus the free energy density. Such a quantity is then quite convenient for understanding the effect of finite size corrections.

\begin{figure}[!ht]\centering\includegraphics[scale=0.25]{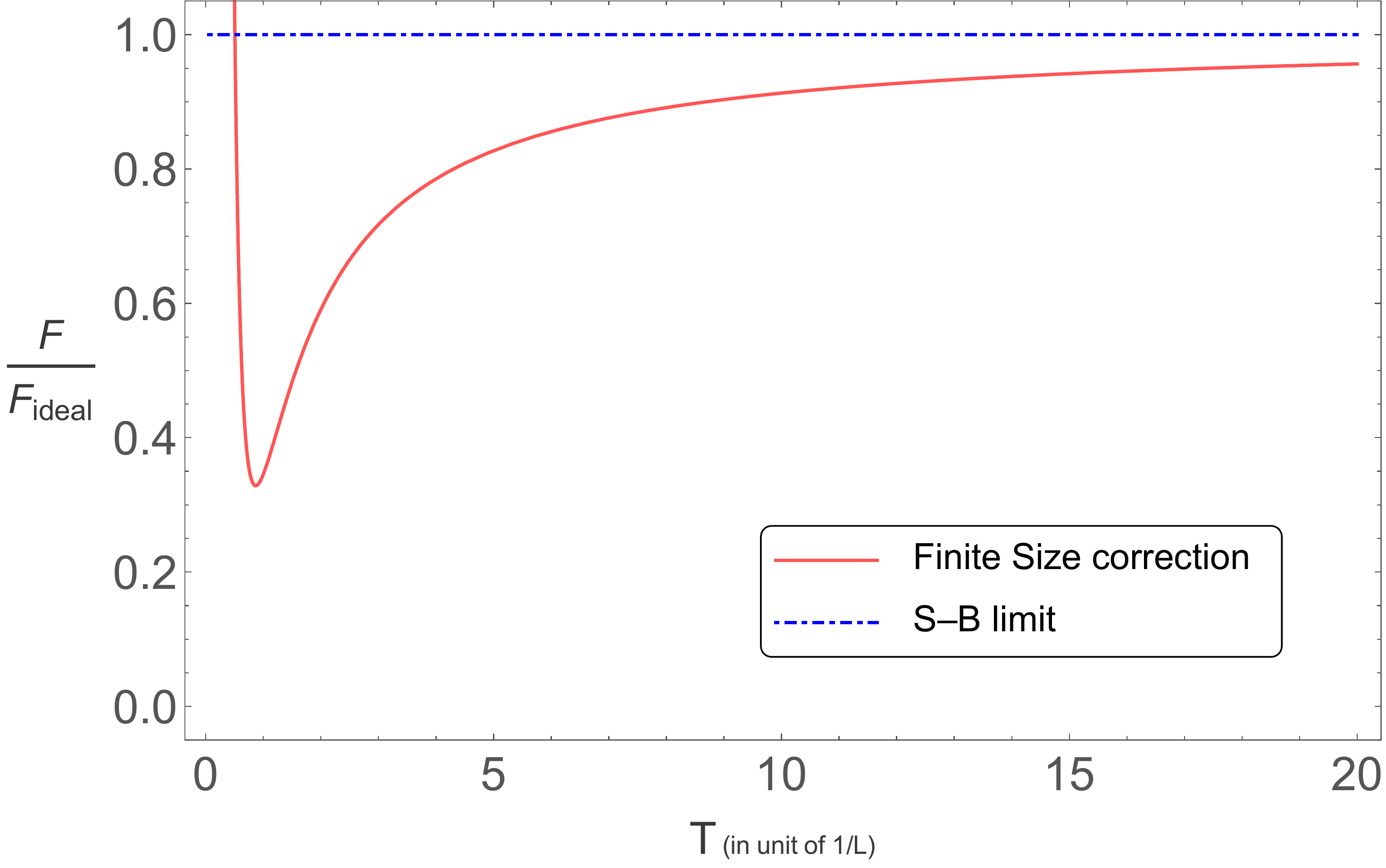}
\caption{Correction to the free energy of a system at temperature $T$ due to the compactification along one direction with length $L$. The result is normalized to its non-compactified, non-interacting limit, and plotted as a function of $T$ in unit of $1/L$.}\label{FreeEnergyFig}
\end{figure}

In Figure~\ref{FreeEnergyFig}, we notice a sharp increase at low temperature which is simply a consequence of the fact that the function is normalized to the fourth power of temperature. Indeed, the zero temperature limit of the free energy is finite at fixed $L$: It is the so-called Casimir value~(\ref{FreeEnergyDensityZeroTemp}).

\subsubsection{Non Additivity of the Equation of State ``Entropy Versus Temperature'' in Finite Size Systems}

We now wish to bring to the attention of the readers that a finite size system may not only be asymmetric, as mentioned previously, but will also lose some of the property of additivity.

\begin{figure}[H]\centering\includegraphics[scale=0.25]{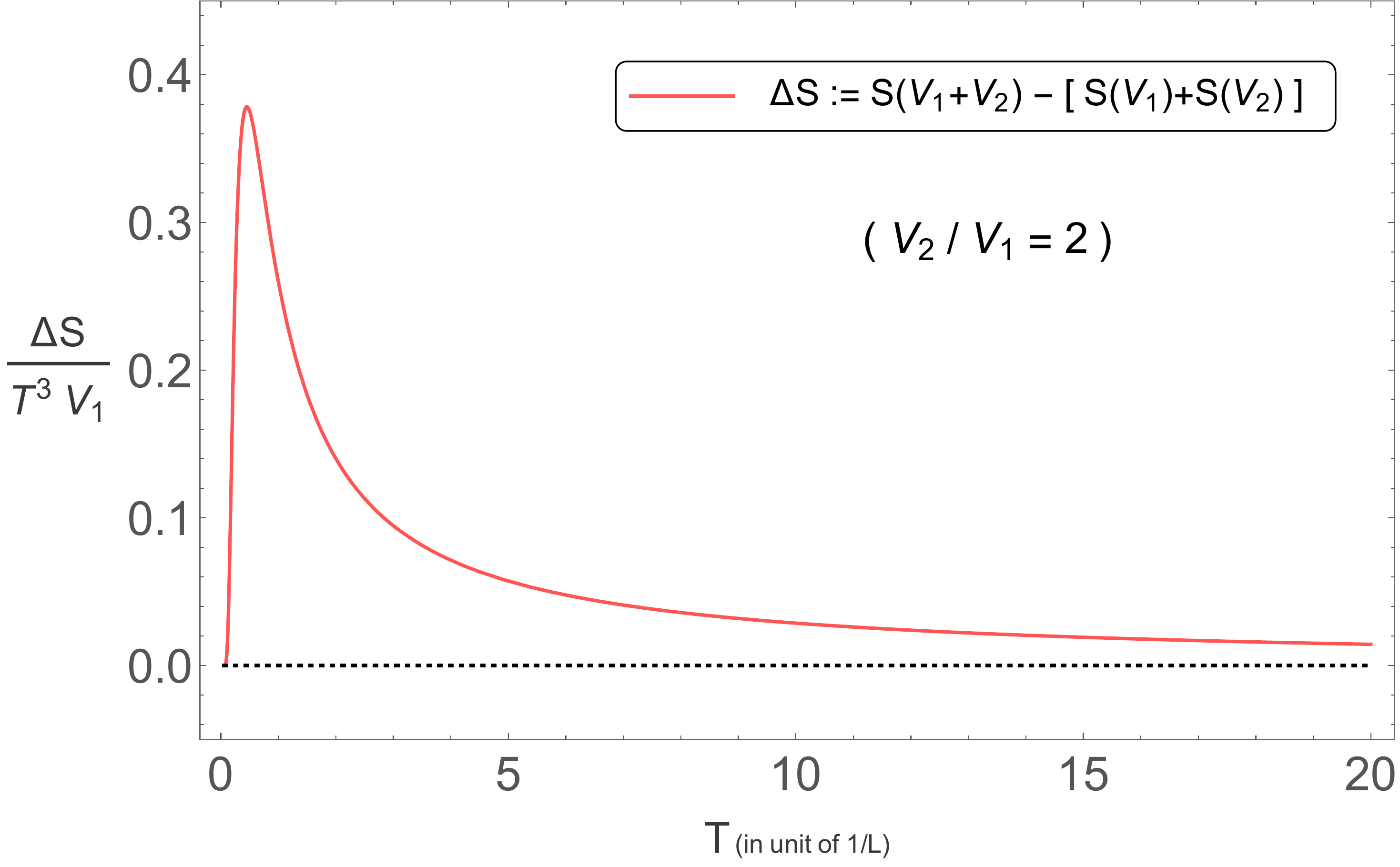}
\caption{Non additivity of the entropy as a function of the temperature, due to the spatial compactification. The quantity is plotted as a function of $T$ in unit of $1/L$, but goes to zero in the limit $L\rightarrow\infty$, and is normalized to appropriate powers of the temperature and volume. One subsystem ($V_2$) is twice as large as the other.}\label{NonAdditivityFig}
\end{figure}

In Figure~\ref{NonAdditivityFig}, we notice that in the large $L$ limit the system becomes fully additive, with for example the equation of state $S(T)$ being additive, as it is expected. However, this happens in the asymptotically small $T$ limit too. But this is merely a consequence of the fact that at zero temperature, the total entropy function vanishes (in the thermodynamic limit, i.e., at present since the volume is infinite, this is called the third law of thermodynamics). We further notice that both subsystems with volumes $V_1$ and $V_2$ have the same temperature, in agreement with the zeroth law of thermodynamics.

\subsubsection{Finite Size Corrections to the Speed of Sound}

Finally, we wish to present the squares of the two possible isochoric speeds of sound in between infinite parallel planes: the one that is transverse to the planes ($c^2_{s_1}$), and the one that is longitudinal ($c^2_{s_{2/3}}$). We also refer to~\cite{To_appear} for much more detail on the derivation of such a quantity. This sound, propagating in an asymmetric manner, can be understood as due to variations in the pressures (both longitudinal and transverse) and the energy density as a consequence of a certain heat transfer from/toward the system.

\begin{figure}[H]\centering\includegraphics[scale=0.235]{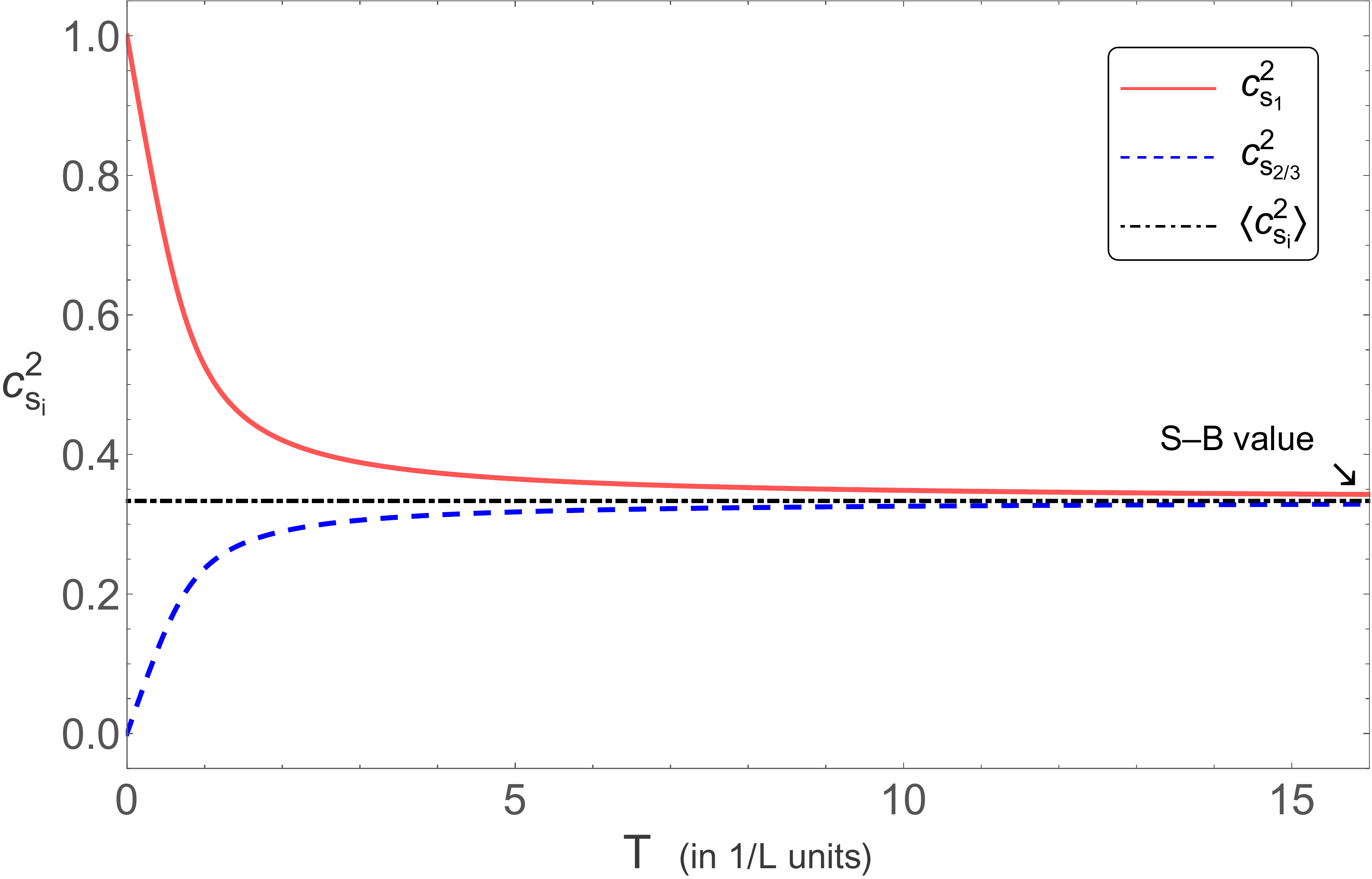}~~\includegraphics[scale=0.24]{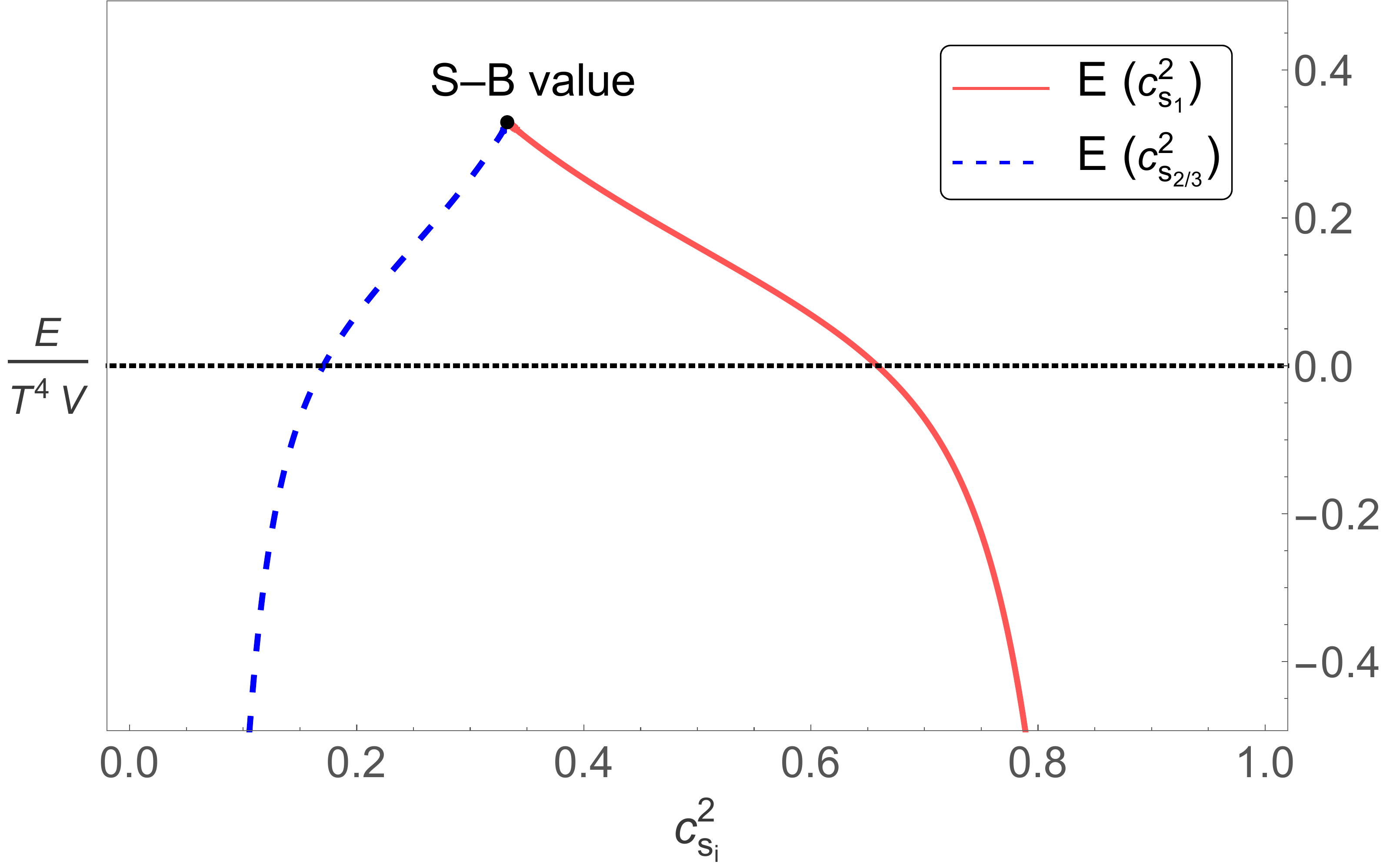}
\caption{\textbf{Left} figure, we display the two isochoric speeds of sound in between infinite parallel planes, as a function of $T$ in unit of $1/L$. \textbf{Right} figure, we display the total energy of the system as a function of each of the isochoric speeds of sound: The transverse one ($c_{s_1}$), and the longitudinal ones ($c_{s_{2/3}}$).}\label{CsFig}
\end{figure}

Both Figures~\ref{CsFig} (left and right) have a number of interesting features, and we also refer to~\cite{To_appear} for a more complete interpretation of the result. However, it is already noteworthy that the average of the isochoric speeds of sound in the three directions ($c_{s_1}$, $c_{s_2}$, and $c_{s_3}$; the black line on the left Figure~\ref{CsFig}) is identically equal to the well known non-compactified limit of $c_s=1/\sqrt{3}$. For both velocities, the correction seems not to be negligible anymore, starting at about 6--10~\% for system sizes relevant to lead--lead collisions at a few hundred MeV. On Figure~\ref{CsFig} (right), we display the total energy of such a system, as a function of both isochoric speeds of sound. The negative region for the total energy is simply the consequence of a certain Casimir effect, notifying the fact that the thermal contribution to the energy is not dominant anymore.


\section{Conclusions}\label{sec:Conclu}

Despite tremendous advances, with various analytic and numerical methods, in the understanding of the thermodynamics of quark--gluon plasmas as created in HIC, we have presented a few preliminary results that suggest the need for a further refinement in the overall picture. This need of improvement seems to be valid, as far as can be understood at present, for quantities related to the thermodynamics of the systems. They seem nevertheless to play an important role in understanding the degrees of freedom at work across the quark--hadron phase transition better. However, the question remains whether or not more dynamical quantities could be less sensitive to the finite size of the system.


\pagebreak
\vspace{6pt}
\acknowledgments{S.M.~would like to acknowledge the financial support from the Claude Leon Foundation, and the South African National Research Foundation (NRF) for supporting his research mobility. S.M.~would like to thank the organizers of ``Compact stars in the QCD Phase Diagram VI'', the Joint Institute for Nuclear Research in Dubna, and in particular David Blaschke and Alexander Sorin. I.K.~and W.A.H.~wish to thank the SA-CERN Collaboration, the NRF, and the South African National Institute for Theoretical Physics for their support. I.~K.~wishes to further acknowledge Deutscher Akademischer Austauschdienst for supporting her work.}


\conflictsofinterest{The authors declare no conflict of interest.}


\reftitle{References}



\begin{thebibliography}{99}
\bibitem{Convergence&WeakCoupling_Blaizot}
Blaizot,~J.P.; Iancu,~E.; Rebhan,~A. On the apparent convergence of perturbative QCD at high temperature. %
{\it{Phys.\ Rev. D}} {\bf{2003}}, \emph{68}, 025011.
\bibitem{Peikert_Tc}
Karsch,~F.; Laermann,~E.; Peikert,~A. Quark mass and flavor dependence of the QCD phase transition. %
{\it{\mbox{Nucl.\ Phys. B}}} {\bf{2001}}, \emph{605}, 579--599. %
\bibitem{Bazavov_Tc}
Bazavov,~A.; Bhattacharya,~T.; Cheng,~M.; DeTar,~C.; Ding,~H.-T.; Steven~Gottlieb; Gupta,~R.; %
Hegde,~P.; \mbox{Heller,~U.M.;} Karsch,~F.; et al. The chiral and deconfinement aspects of the QCD transition. %
{\it{Phys.\ Rev. D}} {\bf{2012}}, \mbox{\emph{85}, 054503}.
\bibitem{deForcrand:2010ys}
De Forcrand,~P. Simulating QCD at finite density. %
{\it{PoS LAT}} {\bf{2009}}, \emph{2009}, 010 %
\bibitem{Gupta:2011ma}
Gupta,~S. QCD at finite density. %
{\it{PoS LAT}} {\bf{2010}}, \emph{2010}, 007 %
\bibitem{Blaizot:2001vr}
Blaizot,~J.P.; Iancu,~E.; Rebhan,~A. Quark number susceptibilities from HTL resummed thermodynamics. %
{\it{\mbox{Phys.\ Lett. B}}} {\bf{2001}}, \emph{523}, 143--150.
\bibitem{Andersen:2012wr}
Andersen,~J.O.; Mogliacci,~S.; Su,~N; Vuorinen,~A. Quark number susceptibilities from resummed perturbation theory. %
{\it{Phys.\ Rev. D}} {\bf{2013}}, \emph{87},  074003.
\bibitem{Mogliacci:2013mca}
Mogliacci,~S.; Andersen,~J.O.; Strickland,~M.; Su,~N; Vuorinen,~A. Equation of State of hot and dense QCD: 
Resummed perturbation theory confronts lattice data. %
{\it{J. High Energy Phys.}} {\bf{2013}}, \emph{1312}, 055.
\bibitem{Haque:2014rua}
Haque,~N.; Bandyopadhyay,~A.; Andersen,~J.O.; Mustafa,~M.G.; Strickland,~M.; Su,~N. %
Three-loop HTLpt thermodynamics at finite temperature and chemical potential. %
{\it{J. High Energy Phys.}} {\bf{2014}}, \emph{1405}, 027.
\bibitem{RHIC}
Tannenbaum,~M.J. Highlights from BNL-RHIC. %
{\it{arXiv}} {\bf{2012}}, arXiv:1201.5900.
\bibitem{LHC}
M\"{u}ller,~B.; Schukraft,~J.; Wyslouch,~B. First Results from Pb+Pb collisions at the LHC. %
{\it{Ann.\ Rev.\ Nucl.\ Part.\ Sci.}} {\bf{2012}}, 62, 361--386.
\bibitem{FAIR}
Heuser,~J.M.~[CBM collaboration]. The compressed baryonic matter experiment at FAIR. %
{\it{Nucl.\ Phys. A}} {\bf{2013}}, \emph{904--905}, 941c--944c.
\bibitem{NICA}
Kekelidze,~V.; Kovalenko,~A.; Lednicky,~R.; Matveev,~V.; Meshkov,~I.; Sorin,~A.; Trubnikov,~G. Project NICA at JINR. %
{\it{Nucl.\ Phys. A}} {\bf{2013}}, \emph{904--905}, 945c--948c.
\bibitem{Mogliacci:2013iwa}
Mogliacci,~S. Kurtoses and high order cumulants: Insights from resummed perturbation theory. %
{\it{J.\ Phys.\ \mbox{Conf.\ Ser.}}} {\bf{2014}}, \emph{503}, 012005.
\bibitem{Aamodt:2011mr}
Aamodt,~K.; Abrahantes~Quintana,~A.; Adamov\'{a},~D.; Adare,~A.M.; Aggarwal,~M.M.; Aglieri~Rinella,~G.; Agocs,~A.G.; Aguilar~Salazar,~S.; %
Ahammed,~Z.; Ahmad,~N.; et al. [ALICE Collaboration]. Two-pion Bose-Einstein correlations in central Pb-Pb collisions at $\sqrt{{s}_{NN}} =$ 2.76 TeV. 
{\it{Phys.\ Lett. B}} {\bf{2011}}, \emph{696}, 328--337.
\bibitem{Adam:2015pya}
Adam,~J.; Adamov\'{a},~D.; Aggarwal,~M.M.; Aglieri~Rinella,~G.; Agnello,~M.; Agrawal,~N.; Ahammed,~Z.; Ahmed,~I.; Ahn,~S.U.; Aimo,~I.; et al.~[ALICE Collaboration]. Two-pion femtoscopy in p-Pb collisions at $\sqrt{s_{\rm NN}}=5.02$ TeV. {\it{Phys.\ Rev. C}} {\bf{2015}}, \emph{91}, 034906.
\bibitem{Mogliacci:2014pxa}
Mogliacci,~S. Probing the Finite Density Equation of State of QCD via Resummed Perturbation Theory. %
{\it{Ph.D.Thesis}}, Bielefeld University, Bielefeld, Germany, 2014.
\bibitem{To_appear}
Mogliacci,~S.; Horowitz,~W.A; Kolb\'e,~I. In preparation.
\bibitem{Vuorinen:2016pwk}
Vuorinen,~A. Quark Matter Equation of State from Perturbative QCD. %
{\it{EPJ\ Web\ Conf.}} {\bf{2017}}, \emph{137,} 09011.
\bibitem{Karsch:2015zna}
Karsch,~F.; Morita,~K.; Redlich,~K. Effects of kinematic cuts on net-electric charge fluctuations. %
{\it{Phys.\ Rev. C}} {\bf{2016}}, \emph{93}, 034907.
\bibitem{Satz_HIC}
Satz,~H. Probing the States of Matter in QCD. %
{\it{Int.\ J.\ Mod.\ Phys. A}} {\bf{2013}}, \emph{28}, 1330043.
\bibitem{Koch_Cumulants}
Koch,~V. Hadronic Fluctuations and Correlations. %
{\it{arXiv}} {\bf{2008}}, arXiv:0810.2520.
\bibitem{DimRedPheno1}
Appelquist,~T.; Pisarski,~R.D. High-Temperature Yang-Mills Theories and Three-Dimensional Quantum Chromodynamics. %
{\it{Phys.\ Rev. D}} {\bf{1981}}, \emph{23}, 2305.
\bibitem{DimRedPheno2}
\textls[-15]{Nadkarni,~S. Dimensional reduction in finite-temperature quantum chromodynamics. %
{\it{Phys.\ Rev. D}} {\bf{1983}}, \emph{27}, 917--931.}
\bibitem{BraatenNieto_EQCD}
Braaten,~E.; Nieto,~A. Effective field theory approach to high temperature thermodynamics. %
{\it{Phys.\ Rev. D}} {\bf{1995}}, \emph {51}, 6990.
\bibitem{Kajantie_EQCD}
Kajantie,~K.; Laine,~M.; Rummukainen,~K.; Shaposhnikov,~M.E. Generic rules for high temperature dimensional reduction %
and their application to the standard model. %
{\it{Nucl.\ Phys. B}} {\bf{1996}}, \emph{458}, 90--136.
\bibitem{Linde_IR_Pb}
Linde,~A.D. Infrared Problem in Thermodynamics of the Yang-Mills Gas. %
{\it{Phys.\ Lett. B}} {\bf{1980}}, 96, 289--292.
\bibitem{Mikko&York_Quark_Mass}
Laine,~M.; Schr\"{o}der,~Y. Quark mass thresholds in QCD thermodynamics. %
{\it{Phys.\ Rev. D}} {\bf{2006}}, \emph{73}, 085009.
\bibitem{Keijo&Mikko&York}
Kajantie,~K.; Laine,~M.; Rummukainen,~K.; Schr\"{o}der,~Y. The Pressure of hot QCD up to $g^6 \ln(1/g)$. %
{\it{\mbox{Phys.\ Rev. D}}} {\bf{2003}}, \emph{67}, 105008.
\bibitem{Aleksi_Pressure}
Vuorinen,~A. The Pressure of QCD at finite temperatures and chemical potentials. %
{\it{Phys.\ Rev. D}} {\bf{2003}}, \mbox{\emph{68}, 054017.}
\bibitem{Opt_Kneur}
Kneur,~J.-L.; Neveu,~A. $\alpha_S$ from $F_\pi$ and Renormalization Group Optimized Perturbation. %
{\it{Phys.\ Rev. D}} {\bf{2013}}, \emph{88}, 074025.
\bibitem{Opt_Peter_SPT}
Karsch,~F.; Patk\'{o}s,~A.; Petreczky,~P. Screened perturbation theory. %
{\it{Phys.\ Lett. B}} {\bf{1997}}, \emph{401}, 69--73.
\bibitem{Frenkel&Taylor_HTL}
Frenkel,~J.; Taylor,~J.C. High Temperature Limit of Thermal QCD. %
{\it{Nucl.\ Phys. B}} {\bf{1990}}, \emph{334}, 199--216.
\bibitem{Braaten&Pisarski_HTL}
Braaten,~E.; Pisarski,~R.D. Soft Amplitudes in Hot Gauge Theories: A General Analysis. %
{\it{Nucl.\ Phys. B}} {\bf{1990}}, \emph{337,} 569--634.
\bibitem{Jens&Mike&BraatenHTLpt}
Andersen,~J.O.; Braaten,~E.; Strickland,~M. Hard thermal loop resummation of the thermodynamics of a hot gluon plasma. %
{\it{Phys.\ Rev. D}} {\bf{2000}}, \emph{61}, 014017.
\bibitem{Jens&Mike&BraatenHTLptbis}
Andersen,~J.O.; Braaten,~E.; Strickland,~M. Hard thermal loop resummation of the free energy of a hot quark-gluon plasma. %
{\it{Phys.\ Rev. D}} {\bf{2000}}, \emph{61}, 074016.
\bibitem{HTLpt_Finite_MU_3Loop1}
Haque,~N.; Andersen,~J.O.; Mustafa,~M.G.; Strickland,~M.; Su,~N. Three-loop HTLpt Pressure and Susceptibilities %
at Finite Temperature and Density. %
{\it{Phys.\ Rev. D}} {\bf{2014}}, \emph{89}, 061701.
\bibitem{Bielefeld_Lattice_HighT}
Bazavov,~A.; Ding,~H.-T.; Hegde,~P.; Karsch,~F.; Miao,~C.; Mukherjee,~S.; Petreczky,~P.; Schmidt,~C.; Velytsky,~A. Quark number susceptibilities at high temperatures. {\it{Phys.\ Rev. D}} {\bf{2013}}, \emph{88}, 094021.
\bibitem{WB_Lattice1}
Bors\'{a}nyi,~S.; Fodor,~Z.; Katz,~S.D.; Krieg,~S.; Ratti,~C.; Szab\'{o},~K.K. Fluctuations of conserved charges at finite temperature from lattice QCD. %
{\it{J. High Energy Phys.}} {\bf{2012}}, \emph{1201}, 138.
\bibitem{Bielefeld_Lattice1}
Schmidt,~C. QCD bulk thermodynamics and conserved charge fluctuations with HISQ fermions. %
{\it{J.\ Phys.\ \mbox{Conf.\ Ser.}}} {\bf{2013}}, \emph{432}, 012013v.
\bibitem{Bielefeld_Lattice2}
Schmidt,~C. Baryon number and charge fluctuations from lattice QCD. %
{\it{Nucl.\ Phys. A}} {\bf{2013}}, \emph{904--905}, 865c--868c.
\bibitem{Bazavov_Lattice}
Bazavov,~A.; Ding,~H.-T.; Hegde,~P.; Kaczmarek,~O.; Karsch,~F.; Laermann,~E.; Maezawa,~Y.; Mukherjee,~S.; Ohno,~H.; Petreczky,~P.; %
et al. Strangeness at High Temperatures: From Hadrons to Quarks. %
{\it{Phys.\ Rev.\ Lett.}} {\bf{2013}}, \emph{111}, 082301.
\bibitem{Wup_Bud_Chi4&Ratio}
Bors\'{a}nyi,~S. Thermodynamics of the QCD transition from lattice. %
{\it{Nucl.\ Phys. A}} {\bf{2013}}, \emph{904--905}, 270c--277c.
\bibitem{Stephanov_Kurtosis}
Stephanov,~M.A. On the sign of kurtosis near the QCD critical point. %
{\it{Phys.\ Rev.\ Lett.}} {\bf{2011}}, \emph{107}, 052301.
\bibitem{Ratio_Wupp_Bud_Baryon}
Bors\'{a}nyi,~S.; Fodor,~Z.; Katz,~S.D.; Krieg,~S.; Ratti,~C.; Szabo,~K.K. Freeze-out parameters: lattice meets experiment. %
{\it{Phys.\ Rev.\ Lett.}} {\bf{2013}}, \emph{111}, 062005.
\end{thebibliography}
\end{document}